\documentclass[useAMS,usenatbib,usegraphicx]{mn2e}

\usepackage{amssymb,amsmath}
\usepackage{mathrsfs}
\usepackage{times}
\usepackage{setspace}

\def\aj{AJ}%
\def\araa{ARA\&A}%
\def\apj{ApJ}%
\def\apjl{ApJ}%
\def\apjs{ApJS}%
\def\aap{A\&A}%
\def\mnras{MNRAS}%
\def\prd{Phys.~Rev.~D}%
\def\pasj{PASJ}%
\def\nat{Nature}%
\def\physrep{Phys.~Rep.}%
\newcommand{\msun}{\ensuremath{M_\odot}}
\newcommand{\beq}{\begin{equation}}
\newcommand{\eeq}{\end{equation}}

\newcommand{\mmax}{\ensuremath{\mathscr{M}_{\rm max}}}
\newcommand{\mmin}{\ensuremath{\mathscr{M}_{\rm min}}}
\newcommand{\data}{\ensuremath{\mathbf{D}}}
\newcommand{\ms}{\ensuremath{\mathscr{M}}}

\newcommand{\intd}{\ensuremath{{\rm d}}}
\newcommand{\ye}{\ensuremath{Y_{\rm e}}}
\newcommand{\na}{\ensuremath{N_{\rm A}}}
\newcommand{\kb}{\ensuremath{k_{\rm B}}}

\newcommand{\mpp}{\ensuremath{\mathscr{P}}}

\begin{document}

\voffset -1.5cm

\title[Neutron star masses as a probe of supernova mechanisms]{The observed neutron star mass distribution as a probe of the supernova explosion mechanism}
\author[Pejcha, Thompson and Kochanek]{Ond\v{r}ej Pejcha$^1$, Todd A. Thompson$^{1,2,3}$ and Christopher S. Kochanek$^{1,2}$
\vspace*{6pt}\\
$^1$Department of Astronomy, The Ohio State University, 140 West 18th Avenue, Columbus, OH 43210, USA \\
$^2$Center for Cosmology and Astroparticle Physics, The Ohio State University, 191 West Woodruff Avenue, Columbus, OH 43210, USA \\
$^3$Alfred P. Sloan Fellow \\ 
  email: pejcha@astronomy.ohio-state.edu}

\maketitle

\begin{abstract}
The observed distribution of neutron star (NS) masses reflects the physics of core-collapse supernova explosions and the structure of the massive stars that produce them at the end of their evolution. We present a Bayesian analysis that directly compares the NS mass distribution observed in double NS systems to theoretical models of NS formation. We find that models with standard binary mass ratio distributions are strongly preferred over independently picking the masses from the initial mass function, although the strength of the inference depends on whether current assumptions for identifying the remnants of the primary and secondary stars are correct. Second, NS formation models with no mass fallback are favored because they reduce the dispersion in NS masses. The double NS system masses thus directly point to the mass coordinate where the supernova explosion was initiated, making them an excellent probe of the supernova explosion mechanism. If we assume no fallback and simply vary the mass coordinate separating the remnant and the supernova ejecta, we find that for solar metallicity stars the explosion most likely develops at the edge of the iron core at a specific entropy of $S/\na \approx 2.8\,\kb$. The primary limitations of our study are the poor knowledge of the supernova explosion mechanism and the lack of broad range of SN model explosions of LMC to solar metallicity.
\end{abstract}

\begin{keywords}
binaries: general --- methods: statistical --- pulsars: individual: PSR J1906+0746 --- stars: neutron --- supernovae: general
\end{keywords}

\section{Introduction}
\label{sec:intro}

At the end of the life of a massive star, the degenerate core made of heavy elements grows by episodes of nuclear shell burning until the Chandrasekhar instability initiates collapse. When the central matter reaches nuclear densities, the equation of state stiffens dramatically as a result of the hard core repulsion of the strong force, and a shock wave is launched back into the supersonically collapsing outer mantle. The shock starts at an approximately constant mass coordinate of about $0.5$ to $0.7\,\msun$, which depends mainly on weak-interaction physics and nuclear interactions, and little on the progenitor structure \citep[e.g.][]{goldreich80,yahil83,bethe90}. However, the mass of the iron core is too big to allow for a prompt explosion \citep[e.g.][]{bethe90} and the initial shock wave stalls due to neutrino losses and photodisentegration of heavy nuclei and turns into a standing accretion shock. This is a robust feature of supernova simulations \citep[see][for a review]{janka07}. The mass of the hot central object, the proto-neutron star (PNS), continues to grow by accretion. The PNS cools by emission of neutrinos that are partially absorbed below the shock and may revive its outward movement to produce a core-collapse supernova \citep[e.g.][]{colgate66,bethe85}. The dense central remnant evolves either into a neutron star (NS), or into a black hole (BH), if the explosion fails \citep[e.g.][]{burrows86,liebendorfer01,heger03,kochanek08,oconnor11}, if mass fallback during the explosion increases the mass above the maximum allowed NS mass \citep[e.g.][]{woosley95,zhang08} or if a phase transition occurs in the cooling NS \citep[e.g.][]{brown94,keil95}.

As the supernova explosion develops within the inner core of the star, the overlying layers of stellar material prevent direct observations of the region aside from neutrinos or gravity waves emitted by very nearby supernovae \citep[e.g.][]{ott04,yuksel07,ott12}. Thus, our understanding of the relevant processes is largely based on numerical simulations, which generally do not produce explosions for progenitors more massive than about $15\,\msun$ \citep[e.g.][]{kitaura06,buras06b,marek09,suwa10,takiwaki12,muller12}. Model explosions fail because the neutrino luminosity never reaches the critical value necessary to turn the quasi-stationary accretion shock into an outgoing explosion in a non-rotating progenitor \citep[e.g.][]{bg93,yamasaki05,yamasaki06,murphy08,nordhaus10,pejcha12,fernandez12}. These models with failed explosions continue to accrete until the formation of a BH \citep[e.g.][]{liebendorfer01,oconnor11}. This ``supernova explosion problem'' has remained unsolved for more than four decades. Present day studies are focused on whether multi-dimensional effects will aid the explosion \citep[e.g.][]{herant92,herant94,bhf95,janka96,fryer00,fryer02,fryer04,ohnishi06,iwakami08,murphy08,marek09,nordhaus10,hanke11,takiwaki12}. As these simulations address primarily non-rotating progenitors, it is worth mentioning that the combination of sufficiently rapid rotation and strong magnetic fields can potentially explode a wider range of progenitor stellar masses \citep[e.g.][]{symbalisty84,akiyama03,thompson05,burrows07,dessart08,dessart12}.

In addition to the mass at core bounce and the mass accreted through the shock before explosion, the final mass of the core-collapse remnant is determined by the amount of fallback during the explosion\footnote{The fraction of the PNS mass lost due to neutrino-driven wind in the explosion is negligible \citep[e.g.][]{thompson01}.}. Because self-consistent supernova simulations do not generically explode, the usual approach for studying fallback, supernova nucleosynthesis, and their observational consequences is to initiate an artificial explosion by depositing enough momentum or energy at a specific mass coordinate of the progenitor to match the final supernova energy or nickel yield \citep{woosley95,thielemann96,timmes96,zhang08,dessart11}. The boundary conditions are adjusted for each progenitor star to give an explosion with the desired properties. The amount of fallback in these models depends on the boundary conditions \citep{macfadyen01} and the structure of the progenitor. Broadly speaking, these simplified models predict that low-metallicity stars explode as supernovae when they are still hot and blue. Their compact envelopes thus generate stronger reverse shocks, more fallback, and higher mass remnants \citep{chevalier89,zhang08}. Due to mass loss, solar-metallicity stars have less massive final envelopes, which generically lead to less fallback, and the strongest reverse shocks are produced at the helium/hydrogen interface \citep{woosley95}. Stars without hydrogen envelopes, such as Wolf-Rayet stars, experience little fallback and thus always produce NSs if the supernova mechanism is successful. In any model, more energetic explosions produce less fallback and smaller remnant masses \citep{zhang08}. In addition to increasing the mass of the remnant, variations in fallback could introduce an element of stochasticity into the NS or BH mass function \citep{ozel12}.

Observational constraints on the causal chain that links massive stars, the supernova mechanism and their remnants are difficult to obtain and have many open problems. Examinations of pre-explosion images of type IIp supernovae\footnote{The plateau in the light curve indicates a presence of thick hydrogen envelope and a red supergiant progenitor \citep[e.g.][]{chevalier76,arnett80}.} suggest that the progenitors have initial masses lower than $16.5 \pm 1.5$\,\msun\ \citep{smarttetal09} and, more generally, there is a dearth of high-mass progenitor stars \citep{kochanek08}. The upper limit on the IIp progenitor mass is surprising, because red supergiants with masses of up to $25\,\msun$ are observed \citep{levesque05} and are thought to explode and produce NSs \citep{heger03}. While the possibility that these stars do not explode at all is intriguing \citep{kochanek08}, the statistical significance of this ``red supergiant problem'' is only $2.4\sigma$ \citep{smarttetal09}. Other explanations involve red supergiants exploding as other types of supernovae due to differences in mass-loss rates \citep{smith09,yoon10,moriya11,georgy12}, or binary evolution \citep{eldridge11,smith11}.

As the remnant mass function encodes information about the structure of the progenitors and the supernova explosion mechanism, considerable attention has been devoted to measuring NS and BH masses \citep{finn94,thorsett99,schwab10,kiziltan10,valentim11,zhang11,ozel12}. The most precise mass measurements come from binary systems where one component is a pulsar and there are accurate measurements of at least two post-Keplerian parameters. A special group of such systems are double NS binaries (DNS), where the first-born NS is observed as a recycled pulsar with a rotation period of $20\,{\rm ms} \lesssim P \lesssim 200$\,ms, and the companion NS is the result of a supernova from the initially less massive secondary star \citep[e.g.][]{bhattacharya91,portegies98,burgay03,lyne04}. Six such systems are currently known, giving $12$ precise mass measurements. The masses of these NSs cluster at $\sim 1.35\,\msun$ with a dispersion of only $\sim 0.06\,\msun$ \citep{kiziltan10,ozel12}, and the mean masses of the pulsars and their companions differ by only $0.03\,\msun$ \citep{ozel12}. Present theories argue for very little mass transfer in these systems once the first NS is formed, so both NS masses basically represent the birth masses of the NSs. The mean observed masses are significantly higher than the Chandrasekhar mass of the pre-collapse core, which may indicate growth by fallback during the supernova \citep{kiziltan10}. The tight mass distribution, however, argues against significant fallback, and the properties of the distribution have also been attributed to the particular evolution history that leads to their formation \citep{ozel12}. \citet{schwab10} proposed that a third of the DNSs formed as a result of electron-capture supernovae as evidenced by their lower masses, with the remaining higher mass systems forming as a result of a Fe-core collapse. 

\citet{kiziltan10} and \citet{ozel12} have carefully fit simple analytic models to the DNS mass distributions. Here, we take the additional step of using the mass distribution of NSs to probe the physics of supernova explosions. We replace the parametric models of observed NS masses used by \citet{kiziltan10} and \citet{ozel12} with predictions of NS masses based on the actual physics of progenitors and supernova explosions so that we can directly constrain, compare and assess the validity of physical models for the explosion. We present a Bayesian formalism that quantitatively compares different predictions for the remnant mass function to the DNS data. We compare different DNS production models using the artificial supernova explosion models of \citet{zhang08} with different explosion energies and different progenitor metallicities. In Section~\ref{sec:model}, we outline our Bayesian framework for comparing the NS production models with the observed data. In Section~\ref{sec:results}, we present our results and outline extensions of our model that may further constrain the underlying physics. In Section~\ref{sec:disc}, we discuss our results and their implications for the supernova explosion mechanism.

\section{Statistical model}
\label{sec:model}

In this Section we outline a general Bayesian statistical model to quantitatively evaluate different hypotheses about the origin of the DNS mass distribution. We choose a Bayesian framework because it easily allows for a simultaneous comparison of multiple models with different numbers of parameters, yields best-fit parameter estimates, and naturally incorporates prior knowledge. We start by outlining a procedure for calculating the posterior probability distributions and then we formulate several hypotheses to be evaluated. We also describe the data and the underlying physical model. A major limitation for these models is the very limited availability of supernova explosion models -- even those employing a simple piston at a fixed composition jump or mass cut -- as a function of mass and metallicity.

\subsection{General considerations}

According to Bayes theorem, the posterior probability of hypothesis $H$ with internal parameters $\btheta$ given data $\data$, $P(\btheta H|\data)$, is equal to the prior probability $P(\btheta H)$ multiplied by the marginal likelihood $P(\data|\btheta H)$ that $\data$ arose from hypothesis $H$,
\beq
P(\btheta H|\data) \propto P(\btheta H) P(\data|\btheta H).
\label{eq:bayes}
\eeq
If the data $\data$ are composed of $N$ individual measurements and the $i$-th measurement is characterized by a probability density in an observed pair of masses $\mathbf{M}\equiv (M_1,M_2)$, $P_i(\data|\mathbf{M})$, then the marginal likelihood of hypothesis $H$ is
\beq
P(\data|\btheta H) = \prod\limits_i^{N} \int P_i(\data|\mathbf{M})P_i(\mathbf{M}|\btheta H)\, \intd \mathbf{M},
\label{eq:double_master}
\eeq
where $P_i(\mathbf{M}|\btheta H)$ is the probability that the given value of $\mathbf{M}$ occurs for the parameters $\btheta$ of $H$ for the $i$-th measurement. For a given hypothesis $H$, different values of the parameters $\btheta$ yield different probabilities of the data $P(\data|\btheta H)$ such that we can determine the ``best-fit'' parameters $\btheta$ and their confidence intervals based on the posterior probability distribution $P(\btheta H|\data)$.

Suppose that we have two hypotheses $H_1$ and $H_2$ parameterized by their individual parameter sets $\btheta_1$ and $\btheta_2$. Which of the two hypotheses better describes the data? Within the framework of Bayesian analysis, the relative ``probability'' of the two hypotheses is given by the Bayes factor 
\beq
B_{12} = \frac{B_1}{B_2} = \frac{\int P(\data|\btheta_1 H_1)P(\btheta_1 H_1)\,\intd \btheta_1}{\int P(\data|\btheta_2 H_2)P(\btheta_2 H_2)\,\intd \btheta_2}.
\label{eq:bayes_factor}
\eeq
Note that only the ratio of $B_1$ and $B_2$, $B_{12}$, has any meaning and that it can be extended to an arbitrary number of hypotheses. Proper calculation of $B_{12}$ also requires that the individual probabilities in Equation~(\ref{eq:double_master}) are properly normalized with $\int\! P_i(\data|\mathbf{M})\,\intd \mathbf{M} \equiv 1$, $\int\! P(\mathbf{M}|\btheta H)\,\intd \mathbf{M} \equiv 1$, and $\int\! P(\btheta H)\,\intd \btheta \equiv 1$ over the relevant ranges of $\mathbf{M}$ and $\btheta$. \citet{jeffreys} groups values of $B_{12}$ in several categories: $B_{12} > 10^{1/2}$ implies that hypothesis $H_1$ is ``substantially'' better than $H_2$. If $B_{12} > 10^2$, then the evidence against $H_2$ and in favor of $H_1$ is decisive. \citet{jeffreys} also gives tables to approximately relate $B_{12}$ as a function of number of the parameters in $\btheta$ to a more commonly used $\chi^2$ difference, specifically $B_{12} \propto \exp (-\Delta\chi^2/2)$.

The Bayesian statistical model we present here is similar to the one developed by \citet{ozel10,ozel12} with a key difference: instead of using a phenomenological description based on a parametric function (in their case, a Gaussian), we will tie the observed NS masses directly to physical calculations of remnant masses based on supernova physics and the progenitor structure. This allows us to quantitatively compare different scenarios for the origin of the NS mass distribution.

\subsection{NSs as members of a binary }

The masses of the two NSs in a binary system are not independent and reflect the binary initial mass distribution, any mass transfer processes that occurred during the system evolution, and the supernova physics. We assume that one observation yields a pair of NS masses of the binary, $\mathbf{M} \equiv (M_1,M_2)$, where $M_1$ is the mass of the recycled pulsar, and $M_2$ is the mass of the companion. In Equation~(\ref{eq:double_master}), $P_i(\data|\mathbf{M})$ is the probability of observing the $i$-th pair of masses 
\beq
P_i(\data|\mathbf{M}) = \mathcal{N}(M_1,\overline{M}_{1,i},\overline{\sigma}_{1,i})\mathcal{N}(M_2,\overline{M}_{2,i},\overline{\sigma}_{2,i}),
\label{eq:double_data}
\eeq
and $(\overline{M}_{1,i}, \overline{M}_{2,i})$, $(\overline{\sigma}_{1,i}, \overline{\sigma}_{2,i})$ are the measured masses and their uncertainties for the $i$-th DNS system. Here, $\mathcal{N}$ are Gaussians defined as
\beq
\mathcal{N}(x,\mu,\sigma) = \frac{1}{\sqrt{2\pi \sigma^2}}\exp \left[-\frac{(x-\mu)^2}{2\sigma^2} \right].
\label{eq:gaussian}
\eeq
Equation~(\ref{eq:double_data}) assumes that there are no correlations between the two NS mass measurements in the binary, although these could be included. More complicated models of the probability densities of the observed masses can be included as well.

An important issue for the calculation of $P_i(\mathbf{M}|\btheta H)$ is the assignment of the DNS components to the original primary and secondary stars in the binary. There are two mutually incompatible possibilities, specifically, either the recycled pulsar came from the primary star and the companion from the secondary, or the reverse. Following \citet{press97}, we account for these two probabilities by expressing $P_i(\mathbf{M}|\mmax)$ as a combination of these two mutually incompatible hypotheses
\begin{eqnarray}
P_i(\mathbf{M}|\btheta H) = \int  P(p_i)  \left[ p_i \mpp(M_1,M_2|\btheta H)\right. + \nonumber \\
		  \left. + (1-p_i)\mpp(M_2,M_1|\btheta H)\right] \, \intd p_i,
\label{eq:assig}
\end{eqnarray}
where $p_i$ is the probability that the recycled pulsar in system $i$ came from the primary star and $P(p_i)$ is the prior on $p_i$. $\mpp(M_A,M_B|\btheta H)$ is the probability density distribution of the pair of remnant masses $(M_A,M_B)$ where $M_A$ corresponds to the remnant mass of the primary star and $M_B$ to the that of the secondary star. For an uniform  prior $P(p_i)$, the final probability is a simple average of the two options. However, the binary evolution models generally require that the recycled pulsar originated from the primary star and hence the prior $P(p_i)$ is strongly peaked at $p_i=1$. Based on the binary evolution models, we choose $P(p_i) = \delta(p_i-1)$ for all systems. In Section~\ref{sec:res_binary}, we investigate the appropriate choice of $P(p_i)$ for the individual systems. 

We consider two forms of the probability $\mpp(M_A,M_B|\btheta H)$. First, as a counterpoint to the more complicated models, we consider an ``independent'' star model where each star is independently drawn from the IMF. This simple model is conceptually similar to the parametric models that assume no correlation between the stars in a binary used by \citet{kiziltan10} and \citet{ozel12}. In this model the primary and secondary probability distributions are independent, $\mpp(M_A,M_B|\btheta H) = \mpp(M_A|\btheta H)\mpp(M_B|\btheta H)$, and $\mpp(M|\btheta H)$ is the probability distribution of remnant masses of a single star 
\beq
\mpp(M|\btheta H) \propto\!\! \int\limits_{\mmin}^{\mmax}\!\!\!\!P(\ms)\, \mathcal{N}[M,M'(\ms),\sigma_{\rm theo}]\, \intd \ms.
\label{eq:single}
\eeq
Here, we assume that NSs are produced by stars with initial masses\footnote{Throughout this paper we denote progenitor masses as $\ms$ and remnant masses as $M$.} $\ms$ between $\mmin$ and $\mmax$, where $\mmin$ is fixed and $\mmax$ is a parameter ($\mmax \in \btheta$). $P(\ms)$ is the probability of progenitor mass $\ms$, which we assume to be a power law, $P(\ms) \propto \ms^{-\alpha}$, with $\alpha = 2.35$ to match \citet{salpeter55}. The function $M'(\ms)$ provides the remnant mass $M'$ for the given progenitor mass $\ms$ (see Section~\ref{sec:implement}). The independent model is symmetric and thus Equation~(\ref{eq:assig}) will give the same marginal likelihood with no dependence on $P(p_i)$.

Second, we consider a genuine binary distribution that is defined as
\begin{eqnarray}
\mpp(M_A,M_B|\btheta H) \propto \!\!\!\!\int\limits_{\mmin}^{\mmax}\!\!\!\!\intd \ms_A\!\!\!\! \int\limits_{\ms_{\rm min}}^{\ms_A}\!\!\!\intd \ms_B  P(\ms_A) \frac{P(q)}{\ms_A} \times \nonumber \\
\times \mathcal{N}[M_A,M'(\ms_A),\sigma_{\rm theo}]\mathcal{N}[M_B,M'(\ms_B),\sigma_{\rm theo}],
\label{eq:double_model}
\end{eqnarray}
where we assume that the primaries are drawn from a Salpeter IMF, $P(\ms_A) \propto \ms_A^{-2.35}$, and the secondaries are drawn from a distribution $P(q)$ of mass ratios $q=\ms_B/\ms_A$. We assume either uniform $P(q)$ for $0.02 \leq q \leq 1.0$, or a population of ``twin'' binaries with half of binaries distributed uniformly in the interval $0.9 \leq q \leq 1$ and the other half uniformly distributed for $0.02 \leq q < 0.9$ \citep{pinsono06,kobulnicky07,kochanek09}. We normalize the mass ratio distribution as $\int_{0.02}^1 P(q)\intd q = 1$ and we drop systems with secondaries with $\ms_B < \mmin$ that would produce NS-WD binaries. Systems with more massive primaries thus produce a higher relative fraction of DNSs. We neglect all binary evolution processes that could modify the relation between the initial and remnant masses $M'(\ms)$, because the mass transfer in a DNS progenitor binary system occurs after the main sequence evolution, which fixes the size of the helium core of the primary \citep[e.g.][]{bhattacharya91,portegies98}. We thus assume that $M'(\ms)$ is the same for primaries and secondaries. Again,  $\mmax$ is a free parameter ($\mmax \in \btheta$).

We chose the form of Equations~(\ref{eq:single})--(\ref{eq:double_model}) for several reasons. First, the function $M'(\ms)$ is usually tabulated only for a discrete set of $\ms$ and we need $\mpp(M|\btheta H)$ and $\mpp(M_A,M_B|\btheta H)$ to be continuous and smooth. This is because the likelihoods of many of the NS mass measurements are sharply peaked and we do not want our results to be sensitive to the exact position of the NS mass with respect to the discrete mass models of the theoretical studies. Second, the width of the kernel, $\sigma_{\rm theo}$, can be interpreted as the uncertainty in the theoretical NS masses either due to progenitor structure, NS growth during the accretion phase, or stochasticity in the amount of fallback. In principle, one can have $\sigma_{\rm theo} = \sigma_{\rm theo}(\ms)$ and make the NS mass uncertainty depend on the progenitor mass. If $\sigma_{\rm theo}$ is too small, $\mpp(M|\btheta H)$ and $\mpp(M_A,M_B|\btheta H)$ will have many individual peaks, while if it is too large, the structure in the NS distribution will be smeared out. We varied $0.01\,\msun \leq \sigma_{\rm theo} \leq 0.05\,\msun$ and found that for higher values of $\sigma_{\rm theo}$ the relative probabilities of the models were smaller. However, the ordering of the models did not change. We choose $\sigma_{\rm theo} = 0.025\,\msun$ as a rather arbitrary compromise between the two extremes. This width is several times smaller than the typical width of the DNS mass distributions of \citet{kiziltan10} and \citet{ozel12}.

The last quantity we need to evaluate Equation~(\ref{eq:bayes}) is the prior on $\mmax$, $P(\mmax)$. Since $\mmax$ attains only positive values and we do not have any physical constraints, we set the prior to be uniform in $\ln \mmax$ for $10\,\msun \leq \mmax \leq 100\,\msun$, where the upper limit corresponds to the approximate maximum mass of a star.

\subsection{Data, underlying models and implementation}
\label{sec:implement}

In order to calculate $P(\btheta H|\data)$ for the independent and binary models, we need the mapping between the initial progenitor mass and the final remnant mass $M'(\ms)$. We use the results of \citet{zhang08} summarized in Table~\ref{tab:zhang}, who obtained NS and BH mass distributions for primordial ($Z=0$) and solar metallicity ($Z=Z_{\sun}$) progenitors by positioning a piston at a particular mass coordinate and injecting enough momentum to obtain an explosion with the desired ejecta kinetic energy $E$ at infinity. The pistons were positioned either at the point where the entropy $S/\na=4\,k_{\rm B}$, which corresponds approximately to the base of the oxygen burning shell, or at the edge of the deleptonized core (``$\ye$ core''), which is located deeper in the star where the electron fraction $\ye$ decreases due to electron captures on protons. This radius roughly corresponds to the iron core. We also consider remnant masses that correspond to the $\ye$ core and $S/\na=4\,\kb$ masses with no fallback. Looking at Table~\ref{tab:zhang}, we see that the model calculations do not extend all the way to the minimum mass for supernova explosion $\mmin$. It is expected that fallback is negligible for these low mass stars and that the remnant mass is equal to the core mass. Following \citet{zhang08}, we extend the properties of the $10\,\msun$ stars down to $\mmin$ for primordial composition stars. For solar metallicity and the piston at $S/\na=4\,k_{\rm B}$, we set $M=1.37\,\msun$ for $11 \leq \ms \leq 12\,\msun$ and $M=1.35\,\msun$ for $9.1 \leq \ms < 11\,\msun$. For solar metallicity and the piston at the $\ye$ core, we set $M=1.32\,\msun$ for $9.1 \leq \ms \leq 12\,\msun$. All remnant masses were corrected for the loss of binding energy $E_{\rm bind}$ due to neutrino emission during the supernova event using the approximation
\beq
E_{\rm bind} = 0.075\,\msun \left(\frac{M_{\rm grav}}{\msun}\right)^2,
\label{eq:grav}
\eeq
where $M_{\rm grav}$ is the gravitational mass of the remnant after the correction for $E_{\rm bind}$ \citep{timmes96}. The commonly assumed value of $0.084\,\msun$ for the leading factor from \citet{lattimer89} differs slightly from Equation~(\ref{eq:grav}). We adopt the sample of DNS from \citet{ozel12}, which is reproduced in Table~\ref{tab:dns} for convenience. The sample consists of $6$ NS binaries, that yield $12$ precise NS mass measurements. 

Because we used Gaussians for $P(\data|\mathbf{M})$ and the kernels appearing in Equations~(\ref{eq:single}) and (\ref{eq:double_model}), we can evaluate the integrals over $\mathbf{M}$ in Equation~(\ref{eq:double_master}) analytically by swapping the order of integration. This greatly speeds up the calculation, especially for the binary models. Integrals over $\ms$ in Equations~(\ref{eq:single}) and (\ref{eq:double_model}) were evaluated using the midpoint rule centered on the given progenitor $\ms$ and with $\intd \ms$ equal to half the distance in $\ms$ to the nearest progenitor models. We did not use a more sophisticated integration method because the NS mass distribution is not intrinsically smooth and because there are significant jumps in $M'$ between progenitors of similar mass. We use $5$ to $10$ points for the mass ranges where no directly calculated progenitors are available ($9.5\leq \ms \leq 10\,\msun$ for $Z=0$ and $9.1\leq \ms \leq 12\,\msun$ for $Z=Z_{\sun}$). We assume that the maximum gravitational NS mass is $2.0\,\msun$ \citep{demorest10}, and we thus do not include in the calculation of $\mpp(M|\btheta H)$ or $\mpp(M_A,M_B|\btheta H)$ any progenitor producing $M'(\ms) > 2.0\,\msun$. These progenitors are assumed to yield BHs.

\section{Results}
\label{sec:results}

We first discuss the DNS mass distribution and the ambiguities in associating a NS in a binary with a progenitor (Section~\ref{sec:res_binary}). Then we examine a range of hypotheses on the origin of the NS mass distribution (Section~\ref{sec:res_bfactor}). We also present several extensions and limitations to our analysis (Section~\ref{sec:res_limit}). 

\subsection{Properties of the binary model}
\label{sec:res_binary}

\begin{figure*}
\centering
\includegraphics[width=0.8\textwidth]{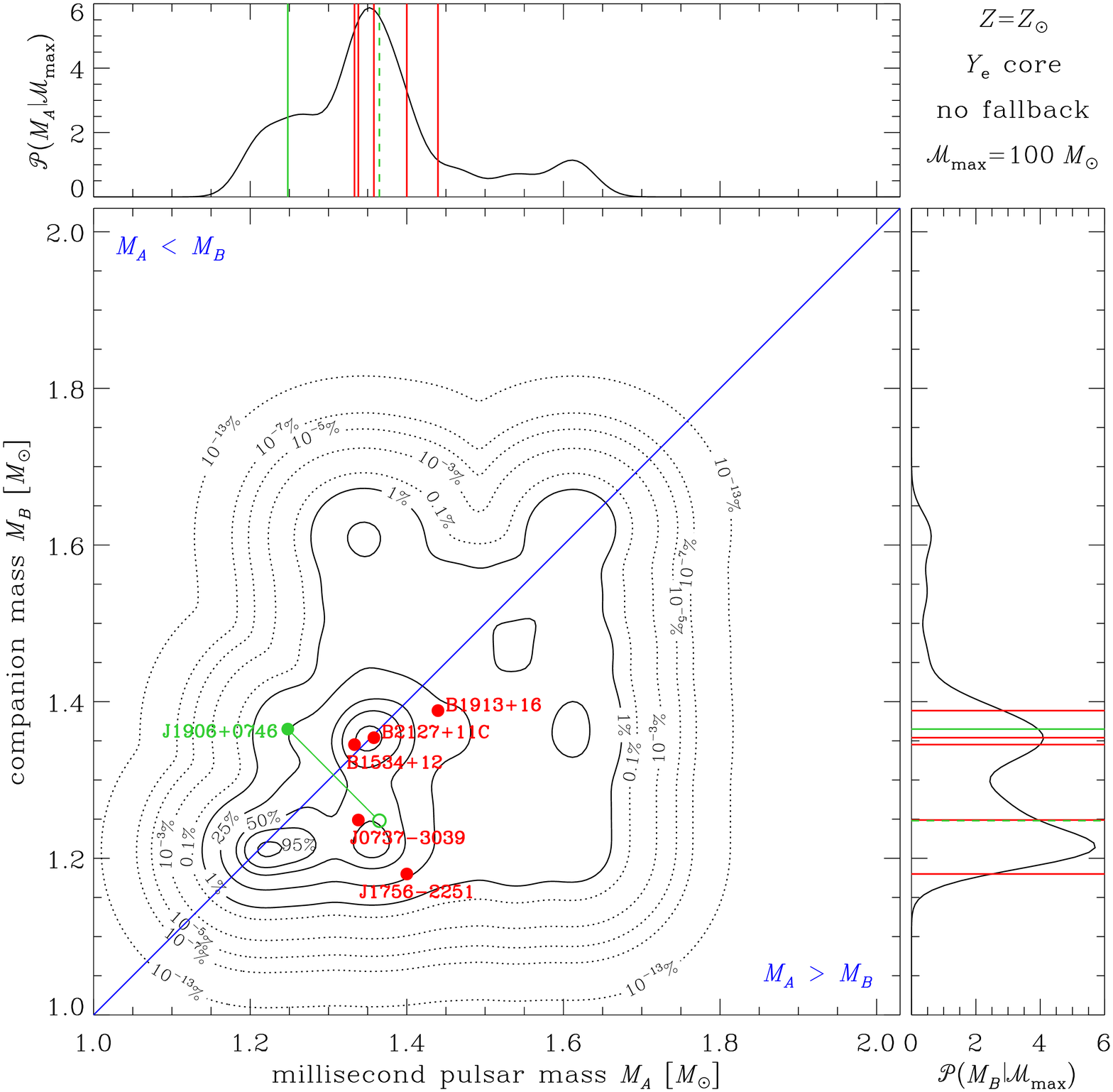}
\caption{Contours of probability $\mpp(M_A,M_B|\mmax)$ for having a DNS with masses $M_A$ and $M_B$ that originated from the more massive primary and less massive secondary stars of the original binary, respectively, for the most probable model discussed in Section~\ref{sec:res_bfactor}. The model has uniform distribution of $q$ and the remaining parameters are given in the upper right corner of the plot. Contour labels indicate the total probability that lies {\em outside} the contour (i.e.\ the contour labeled as ``$0.1\%$'' encloses $99.9\%$ of the total probability). Circles indicate pairs of DNS masses. For the red systems, the pulsar is always assumed to arise from the primary. The green circle is for J1906$+$0746 where we show both possible assignments (see text). The upper panel shows the distribution of remnant masses $M_A$ from the primary, with the lines marking the observed values of $\overline{M}_1$. For J1906$+$0746, we show both cases, where the pulsar came from the primary (solid) or the secondary (dashed). The right panel shows the distribution in $M_B$.}
\label{fig:binary2d}
\end{figure*}

\begin{figure*}
\centering
\includegraphics[width=0.8\textwidth]{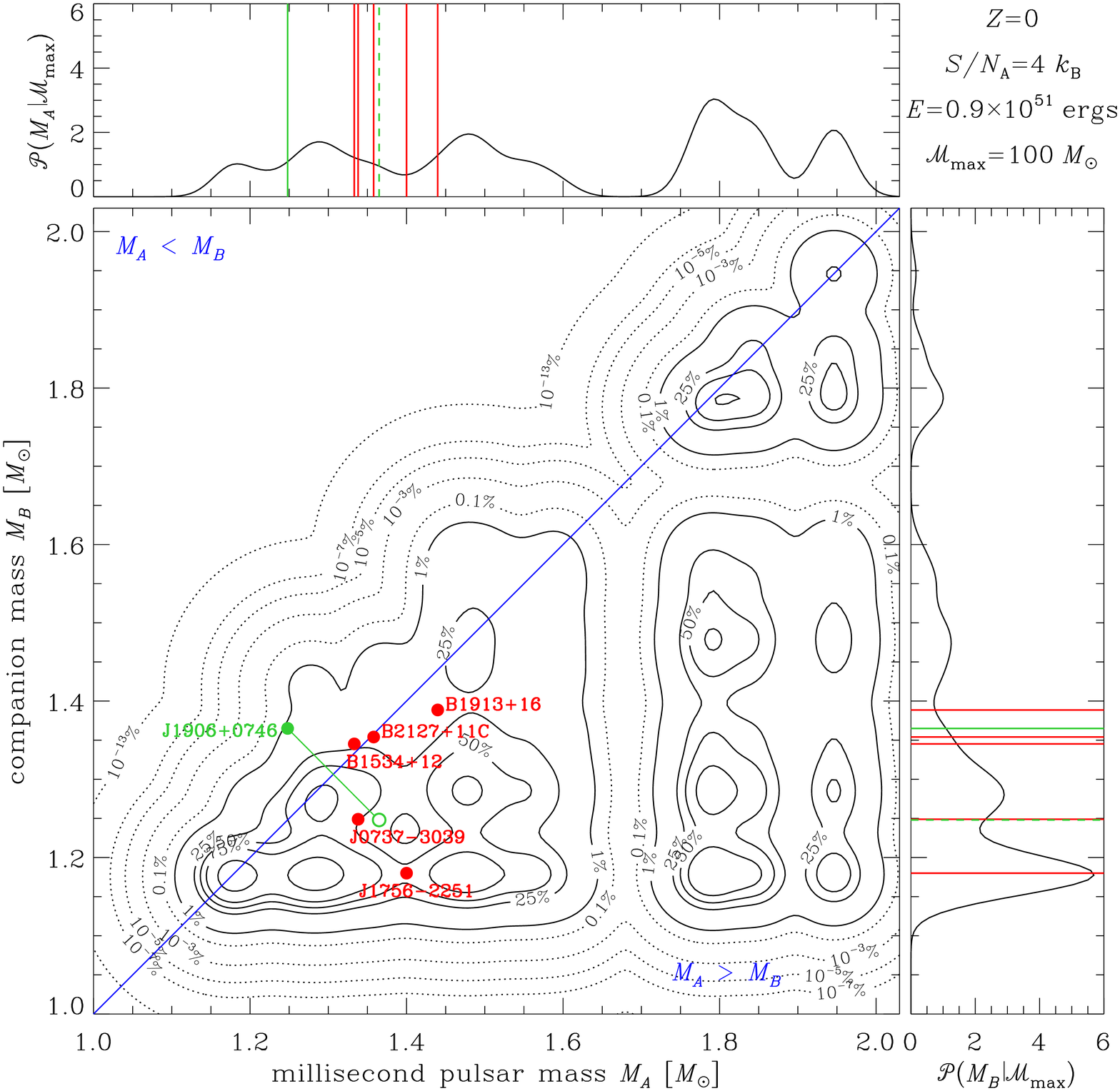}
\caption{Same as in Figure~\ref{fig:binary2d}, but for a model with significant mass fallback. The model has uniform $P(q)$ and the remaining parameters are given in the upper right corner of the plot. The relative probability $B_{12}$ of this model is a factor of $\sim 10^{2.8}$ lower than for the model in Figure~\ref{fig:binary2d} (see Section~\ref{sec:res_bfactor} for more details). }
\label{fig:binary2d_fallback}
\end{figure*}

In Figures~\ref{fig:binary2d} and \ref{fig:binary2d_fallback}, we show examples of the probability distribution of NS masses originating from a binary $\mpp(M_A,M_B|\mmax)$ with a uniform $P(q)$ along with the distributions marginalized over $M_A$ or $M_B$. The model in Figure~\ref{fig:binary2d} has the highest relative probability of all models considered in Section~\ref{sec:res_bfactor}. For comparison, Figure~\ref{fig:binary2d_fallback} shows a model with fallback that has relative probability $B_{12}$ (Eq.~[\ref{eq:bayes_factor}])  lower by a factor of $\sim 10^{2.8}$. Examining Equation~(\ref{eq:double_model}), we see that the primary and secondary mass ranges producing NS are quite different. For uniform $P(q)$, the probability distribution of primary masses $\ms_A$ is proportional to $\ms_A^{-\alpha}(1 - \mmin/\ms_A)$, and primaries with masses close to $\mmin$ do not contribute to the distribution of NS masses $M_A$ because these systems mostly produce NS-WD binaries. The primary distribution producing NSs peaks at $(\alpha+1)\mmin/\alpha \approx 1.43 \mmin$ for a Salpeter IMF. The distribution of secondaries producing NSs is proportional to $\ms_B^{-\alpha} - \mmax^{-\alpha}$ for a flat $P(q)$. Here, the cutoff is at high masses, while the distribution of the lowest mass progenitors is almost Salpeter. These analytic estimates immediately show that NS masses originating from the primary and secondary stars of a binary represent different progenitor mass ranges. For example, for $\mmin=9.1\,\msun$ and $\mmax=25\,\msun$, the mean progenitor masses of the primary and secondary components are $16.5$ and $12.8\,\msun$. This explains why the marginal distribution of $M_A$ in Figure~\ref{fig:binary2d} is not peaked at low $M_A$ and has a stronger secondary peak at $M_A \sim 1.6\,\msun$, when compared to the distribution of $M_B$. The secondary peak is much higher for models that include fallback (Fig.~\ref{fig:binary2d_fallback}). There are no observed DNSs with masses in this second peak, which leads to a preference for models with no fallback (Section~\ref{sec:res_bfactor}). The secondaries $\ms_B$ have progenitor distributions close to the IMF. Thus, the highest peak for secondaries is at $1.22\,\msun$, which is the assumed gravitational NS mass for stars with $9.1 \leq \ms \leq 12\,\msun$.

\begin{figure*}
\centering
\includegraphics[width=0.8\textwidth]{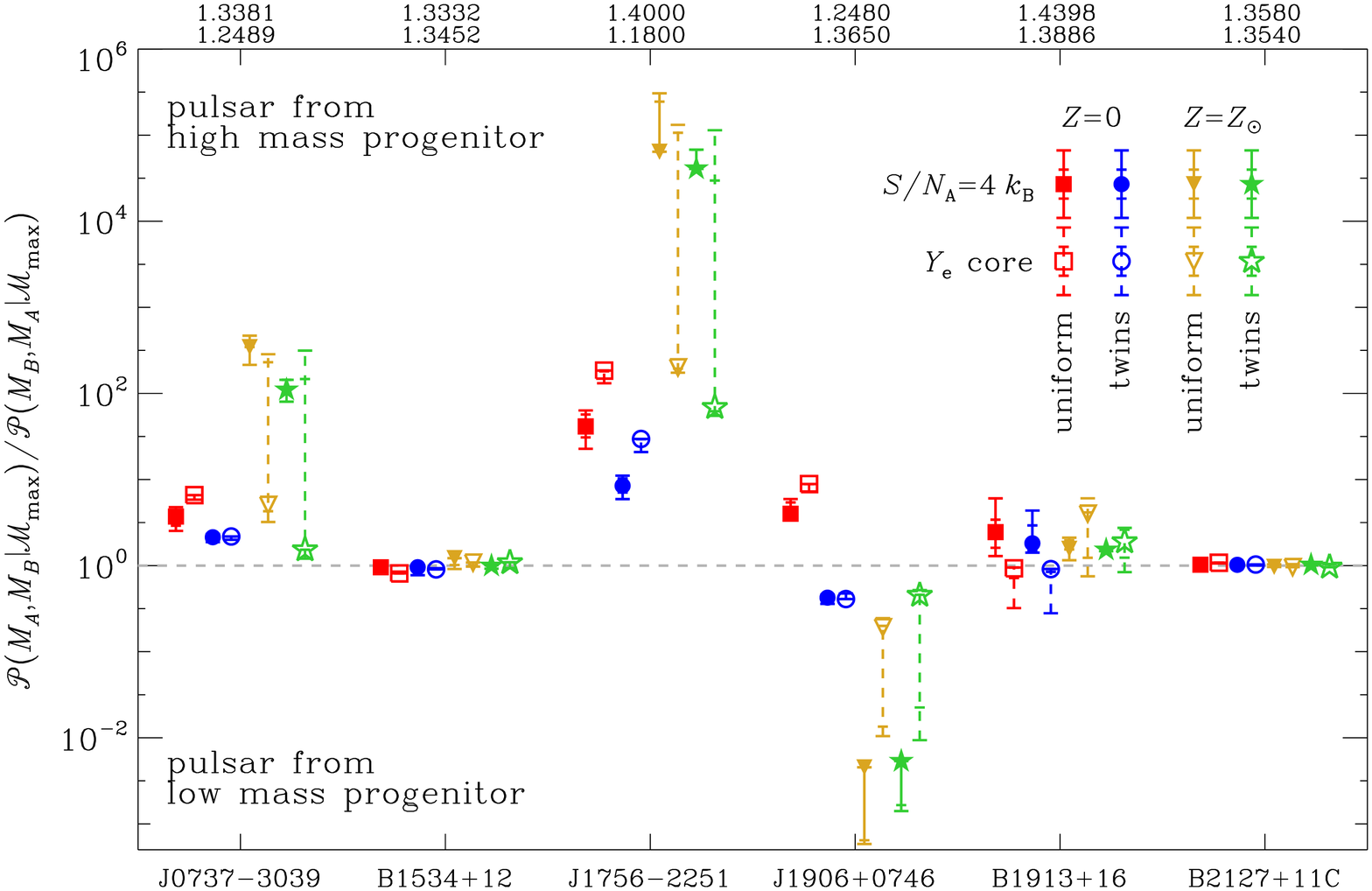}
\caption{Probability that the pulsar in the DNS came from the initially more massive progenitor with respect to the reverse. DNS system names are given at the bottom while the individual NS masses are given at the top with the pulsar masses above the companion masses. The symbols indicate the median, while the error bars shows $1$ and $2\sigma$ contours. The results are shown for the two piston positions, two metallicities and the two binary models (symbols are explained in the plot), and are marginalized over all values of $E$ and $\mmax$, weighted by $P(\mmax|\data)$. Filled squares and circles mark the median, while the error bars show $1$ and $2\sigma$ quantiles.} 
\label{fig:mass_ratios}
\end{figure*}

We see from Figures~\ref{fig:binary2d} and \ref{fig:binary2d_fallback} that most of the probability is in the region where the primary produces a more massive NS. However, since $M'(\ms)$ is not monotonic \citep{zhang08}, there is a small probability that the more massive NS originated in fact from the less massive progenitor. Generally, the farther the NS mass pair is from the diagonal ($M_A=M_B$), the smaller the probability that the more massive NS came from the less massive progenitor. In our formalism we can estimate whether the mass difference between the two NSs is enough to distinguish between an NS originating from the primary or the secondary, or essentially whether the prior $P(p_i) = \delta(p_i-1)$ in Equation~(\ref{eq:assig}) is appropriate. Figure~\ref{fig:mass_ratios} shows the ratio of probabilities that the millisecond pulsar originated from the more massive progenitor, $\mpp(M_1,M_2|\mmax)$, as compared to the reverse $\mpp(M_2,M_1|\mmax)$. In Figure~\ref{fig:mass_ratios} we have marginalized over explosion energies and $\mmax$. We see that if $|M_1-M_2| \lesssim 2\sigma_{\rm theo}$, our model cannot distinguish between the primary/secondary origin of the millisecond pulsar based on the masses alone. The case of J1906$+$0746 is peculiar (because the pulsar is significantly less massive than the companion) and our results show that it is unlikely to have originated from the more massive progenitor. In agreement with \citet{lorimer06} who give a very small characteristic pulsar age (see also our Table~\ref{tab:dns}), we propose that the observed pulsar in J1906$+$0746 comes from the less massive secondary star and we set the prior on $p_i$ in Equation~(\ref{eq:assig}) to be $P(p_i) = \delta(p_i)$ for this system. We show this alternative assignment as an open circle and the dashed lines in Figure~\ref{fig:binary2d}. This alternative assignment increases the relative probability of the binary models by a factor of $\sim 3$ to $\sim 300$.

\subsection{Comparison of the individual models}
\label{sec:res_bfactor}

Next we evaluate the relative probabilities of individual models. We specifically discuss the differences between the independent and binary models, the explosion energy, and the position of the piston. We find that there is little difference in relative probability between the uniform and twin mass ratio distributions so we only discuss the uniform $P(q)$ model, which has slightly higher probability for solar metallicity. By comparing the relative probabilities of models with free $\mmax$ to models that include all progenitors (equivalent to setting $P(\mmax) = \delta(\mmax - 100\,\msun)$ in Eq.~[\ref{eq:bayes_factor}]), we also find that models with $\mmax$ as a free parameter are not significantly preferred. The inferred values of $\mmax$ range from $14\,\msun$ to $35\,\msun$ depending on the method used to infer the ``best-fit'' value, but with confidence intervals covering most of the allowed range for $\mmax$.
Here, we show models marginalized over $\mmax$, although models simply fixing $\mmax=100\,\msun$ give essentially the same results. Finally, for the purposes of this Section, we do the calculations with the pulsar in J1906$+$0746 attributed to the secondary star, consistent with \citet{lorimer06} and our discussion in Section~\ref{sec:res_binary}.

\begin{figure*}
\centering
\includegraphics[width=0.8\textwidth]{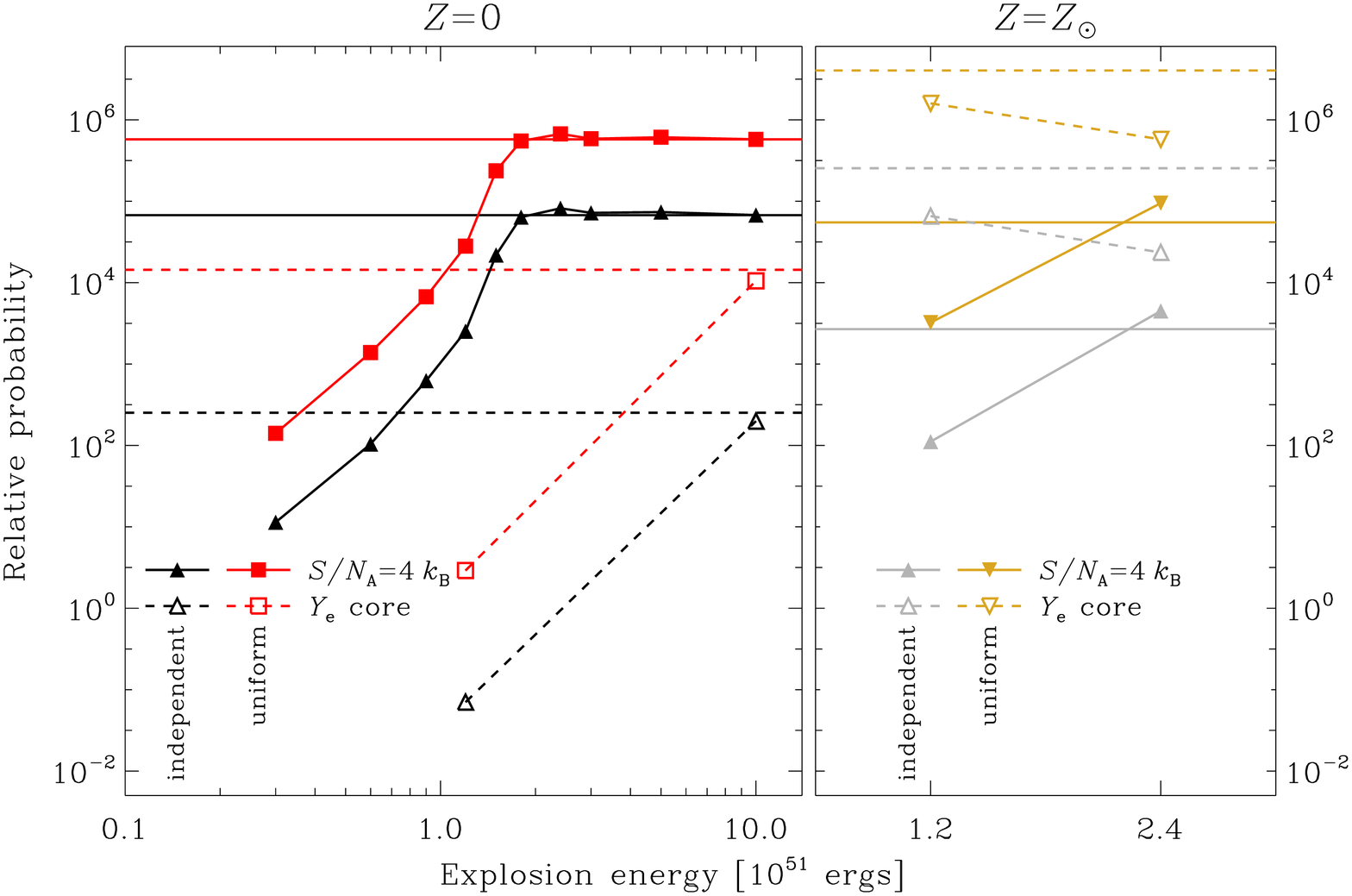}
\caption{Relative probability of the different models for the origin of the DNS mass distribution as a function of the explosion energy and marginalized over $\mmax$. We show primordial composition (left panel) and solar metallicity (right panel) models for the piston positioned at $S/\na = 4\,\kb$ (solid lines with filled symbols) and at the $\ye$ core (dashed lines with open symbols). For the sake of clarity, we show only the independent mass model and the uniform $P(q)$ binary model. The horizontal lines indicate results for the models with NS masses equal to the mass at the piston position and no fallback.}
\label{fig:bayes_factor}
\end{figure*}

Figure~\ref{fig:bayes_factor} shows the relative probabilities (Bayes factors) of our models as a function of explosion energy. Focusing first on the primordial composition models (left panel of Fig.~\ref{fig:bayes_factor}), we see a general trend of increasing model probability for explosion energies up to about $E \approx 2\times 10^{51}$\,ergs, after which the relative probability is essentially constant. The explanation is that higher $E$ explosions produce less fallback and hence reduce the number of high-mass NSs that expand the overall NS mass range. This is confirmed by the models that include only the mass of the core with no fallback (horizontal lines in Figure~\ref{fig:bayes_factor}) that match the probabilities of the high-$E$ models. For solar composition (right panel of Fig.~\ref{fig:bayes_factor}), the effect of $E$ is not clear. 
Since \citet{zhang08} investigated only two values of $E$, the probability ratios are not large (factor of $\sim 3$). 

In all cases, the relative probabilities for binary versus independent mass models is between $5$ and $200$, which suggests a strong preference for binary models unless we assume that the pulsar in J1906$+$0746 came from the initially more massive star. In this case, the relative probability of the binary models decreases to a factor of $\sim 3$ for $Z=0$ and for $Z=Z_{\sun}$ some of the models even disfavor the binary models. This relative change was expected based on Figure~\ref{fig:mass_ratios}. If we assume uniform $P(p_i)$ in Equation~(\ref{eq:assig}), the relative probability of binary models is again only a factor of $\sim 3$ higher than for the independent mass model.  
Binary models are significantly punished if there is a system with $M_A$ significantly lighter than $M_B$ if that is not allowed by the underlying remnant mass model. Correct treatment of the primary/secondary assignment of the millisecond pulsar and companion is crucial for properly evaluating the relative probabilities of the independent and binary models.

Figure~\ref{fig:bayes_factor} also shows clear differences in the relative probabilities of the different piston positions. For $Z=0$, the pistons at the $\ye$ core are strongly disfavored, because for low-mass progenitors the masses of the $\ye$ cores are too low. For solar composition, the situation is reversed. Models with the piston at the $\ye$ core are significantly more likely than those putting it at $S/\na = 4\,\kb$, because they reduce the number of NSs with $M>1.5\,\msun$. Furthermore, we see that the highest probability models essentially correspond to those with no fallback.
There are small changes in the relative probabilities between the two binary mass ratio distributions ($P(q)$) and the primary/secondary assignment for J1906$+$0746. But the best overall model is the one with no fallback, uniform $P(q)$, and remnant masses equal to the $\ye$ core mass. The two piston positions used by \citet{zhang08} are somewhat arbitrary, so in Section~\ref{sec:disc} we address the question of whether some other piston position would produce better agreement with the observations.

\begin{figure*}
\centering
\includegraphics[width=0.78\textwidth]{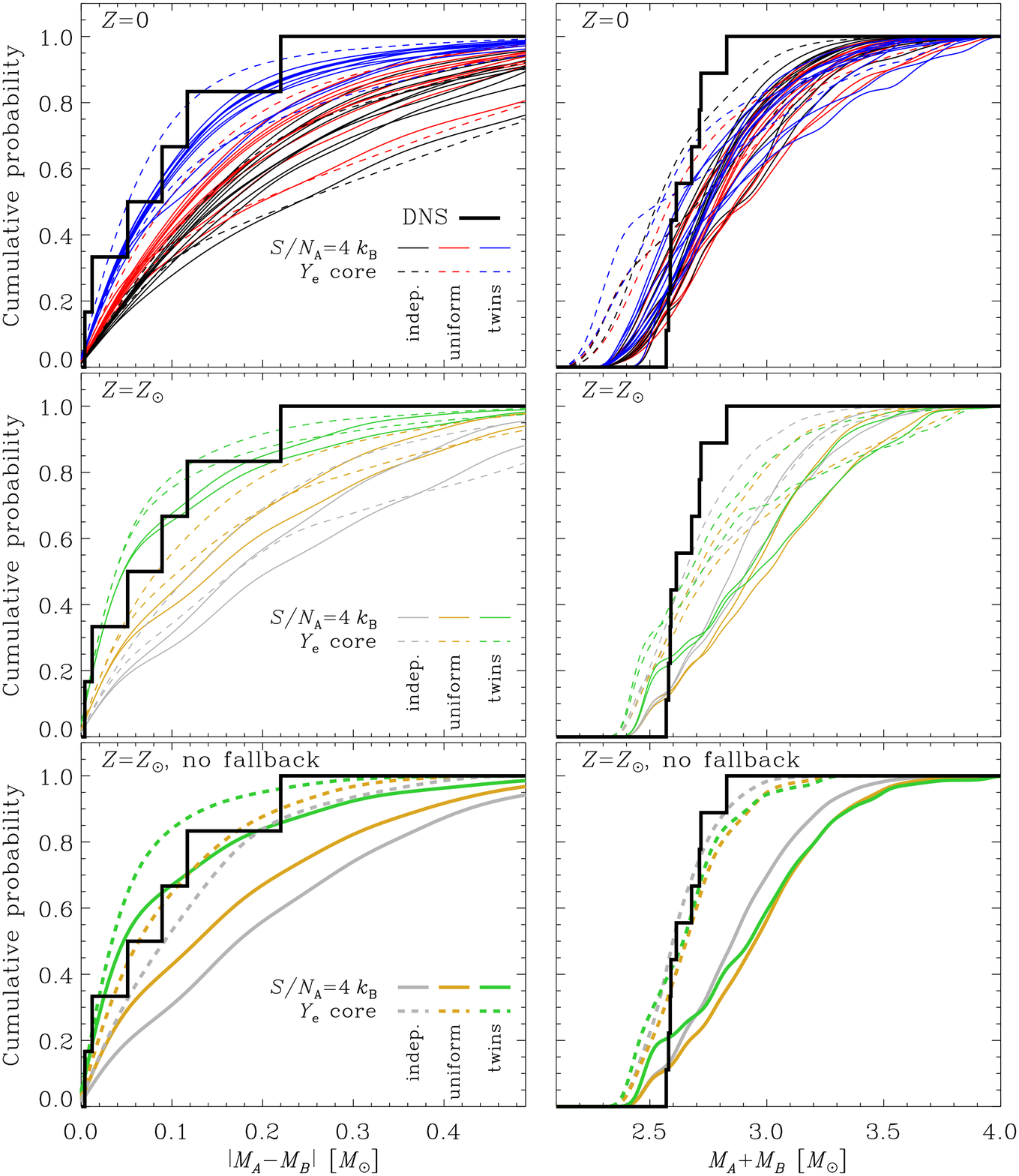}
\caption{Cumulative probability distributions of the mass differences ($|M_A-M_B|$, left panels) and the total masses ($M_A+M_B$, right panels) of the DNS systems. The results of the \citet{zhang08} models with $\mmax=100\,\msun$ are shown with thin lines for $Z=0$ (top panels) and $Z=Z_{\sun}$ (middle panels). The bottom panels show solar metallicity models with no fallback. The observed cumulative distributions of $|\overline{M}_1-\overline{M}_2|$ and $\overline{M}_1+\overline{M}_2$ are shown with the thick black line. The distribution of total masses in the right panel includes additionally DNS systems J1518$+$4904 \citep{janssen08}, J1811$-$1736 \citep{corongiu07}, and J1829$+$2456 \citep{champion05} that have accurate total (but not individual) masses.} 
\label{fig:ns_cumul}
\end{figure*}

While in the rest of this paper we examine the relative probability of different hypotheses for the DNS mass distribution in the Bayesian sense, at this point we make a ``frequentist'' diversion and compare the observations to the models in an absolute sense. We have argued that the individual NS masses in a DNS system reflect the binarity of the original system in the sense that the primary and secondary progenitor and NS mass distributions are different. Even though one of the components is observed as a pulsar, the ambiguity of which component originated from the primary or secondary star of the binary remains present for some systems unless more information is added (J1906$+$0746). In order to compare the models globally, we construct in Figure~\ref{fig:ns_cumul} cumulative distributions of the DNS total mass ($M_1+M_2$) and the mass difference ($|M_1-M_2|$), along with the predictions from \citet{zhang08} models coupled to the scenarios described in Section~\ref{sec:model} with $\mmax=100\,\msun$. 

The distribution of mass differences in the left column of Figure~\ref{fig:ns_cumul} shows that getting the observed mass differences is entirely within the range of the models, although most of them predict a much broader distribution of mass differences. The twin models (blue and green lines) generally give smaller mass differences than their uniform $P(q)$ counterparts. Note that the no fallback, $\ye$ core, solar metallicity model with the highest Bayesian relative probability (thick yellow dashed line in the lower left panel) shows almost perfect agreement with the observations. We see in the cumulative distribution of total masses in the right column of Figure~\ref{fig:ns_cumul} that the observed distribution is much narrower than all theoretical predictions. The narrowest cumulative distributions are again produced by solar metallicity models with no fallback and remnant masses given by the $\ye$ core masses (thick dashed lines). Implementing a cutoff for progenitor mass $\mmax$ makes the distributions narrower. However, this is only weakly favored by the observations. Additionally, the observed minimum total mass of $\sim 2.6\,\msun$ is markedly higher than any of the minimum total masses predicted by the models ($2.2$ to $2.4\,\msun$). The shift is $0.2$ to $0.3\,\msun$, which is much higher than the mass necessary to recycle the pulsar to the observed spin periods. 

At this point in a ``frequentist'' analysis, we would compare the two cumulative distributions using a Kolmogorov-Smirnov test to ascertain whether the observations are compatible with the models in an absolute sense. However, mass differences and total masses are only particular aspects of the full 2D distributions. If we examine Figure~\ref{fig:binary2d}, which shows $\mpp(M_1,M_2|\mmax=100\,\msun)$ for the model with the highest Bayesian relative probability, we see that all $6$ binaries lie within the probability contour containing $75\%$ of the probability (if we assume that the pulsar in J1906$+$0746 came from the initially less massive progenitor). The chance of having no system outside this contour is $0.75^6 = 0.178$ and we would typically expect $4.5 \pm 1.1$ systems within this contour given $6$ systems. This suggests that the highest Bayesian probability model represents the data quite well -- the fact that there is no DNS system with $M > 1.5\,\msun$ is likely only a statistical fluctuation. However, none of the three additional DNS systems with accurate total masses (Fig.~\ref{fig:ns_cumul}, right panels) has a total mass of about $2.4\,\msun$, which suggests that the finer features of the DNS mass distribution might not be entirely reflected in the theoretical models. More DNS systems with accurate masses of both components are necessary to address this question.

\subsection{Extensions and limitations}
\label{sec:res_limit}

There are numerous possible extensions to the analyses presented in this paper. For example, if $\mmin$ is slightly lower, as suggested by studies of Type IIp supernova progenitors \citep{smartt09}, stars in this mass range will dominate the total probability due to the steepness of the Salpeter IMF. Core masses of the progenitors that were not explicitly calculated ($\ms < 10\,\msun$ for $Z=0$ and $\ms < 12\,\msun$ for $Z=Z_{\sun}$) can also be different -- \citet{nomoto84} calculated presupernova structure of a $8.8\,\msun$ star, which has enclosed mass at both $S/\na = 4\,\kb$ and $\ye = 0.499$ of $1.49\,\msun$ (baryonic)\footnote{Note that in this model $S/\na < 4\,\kb$ for all zones below $1.49\,\msun$ except a single zone at $1.19\,\msun$, where it reaches $S/\na = 4.02\,\kb$.}, which is significantly higher than that assumed for low-mass progenitors in the \citet{zhang08} models. If we vary $\mmin \geq 5\,\msun$ and the baryonic remnant mass of $1.2 \leq M_{\rm ec} \leq 1.55\,\msun$ for stars where the progenitor structure was not explicitly calculated, use uniform priors in $\ln \mmin$ and $\ln M_{\rm ec}$, and hold $\mmax = 100\,\msun$ fixed, we find very little change in the relative probabilities of the models. There is no significant change of $M_{\rm ec}$ with respect to \citet{zhang08}. The results on the minimum progenitor mass are typically $\mmin = 8.9^{+1.4}_{-1.5}\,\msun$, which is compatible with both \citet{zhang08} and \citet{smartt09}. We also find $1.32 \lesssim M_{\rm ec} \lesssim 1.39\,\msun$ for $Z=Z_{\sun}$, which is compatible with \citet{zhang08}.

The current DNS sample has only six systems with precise masses. A precise measurement of a NS birth mass greater than about $1.5\,\msun$ would greatly constrain the NS formation models. Until such system is found\footnote{NS masses greater than $1.5\,\msun$ have been measured. For example, the system J1614$-$2230 has $M=1.97 \pm 0.04\,\msun$ \citep{demorest10}. However, these NSs are likely significantly recycled and do not represent NS birth masses, although \citet{lin11} and \citet{tauris11} argue that the birth mass of J1614$-$2230 was higher than $1.5\,\msun$.}, it might be possible to obtain additional constraints by including NS systems with less precise mass measurements. The Bayesian formalism naturally accounts for non-Gaussian marginal likelihoods $P_i(\data|M)$ such as those presented by \citet{ozel12} for some systems. Adding DNS systems with only a precise total mass measurement is unlikely to change our results, as their total masses are compatible with our sample (Fig.~\ref{fig:ns_cumul}). We experimented with adding the eclipsing X-ray pulsars, which should also have masses near the birth mass \citep{rawls11, ozel12}, and found that these measurements have uncertainties that are too large to improve the constraints.

The models can also be extended to include other types of binaries with degenerate components (e.g.\ BH-NS, BH-BH, NS-WD) since we can calculate the full remnant mass function for the binaries (Eq.~[\ref{eq:double_model}]). That there are no known BH-NS binaries must strongly constrain $\mmax$ through the relative probabilities of NS-NS, NS-BH and BH-BH for different values of $\mmax$. Unfortunately, this also requires estimates for the relative detection efficiencies of the individual channels.

Finally, there are also a number of limitations to this work. The current sample of DNSs with precise masses has only six systems. The best available remnant mass function of \citet{zhang08} is based on 1D models that artificially explode non-rotating progenitors produced by a single stellar evolution code \citep{woosley02}. In addition, specific conditions have to be met to produce a DNS system. \citet{belczynski02} gives a comprehensive list of possible channels for DNS formation. Most of them involve mass transfer and a theoretically uncertain phase of common envelope evolution, which exposes the NS to potentially hypercritical accretion \citep[e.g.][]{chevalier93,brown95,dewi06,lombardi11}. Furthermore, \citet{belczynski10} investigated the relative numbers of DNS systems and isolated recycled pulsars and found a disagreement with theoretical predictions that point to a lack of understanding of massive binary star evolution or supernova explosions.
Another significant effect is the disruption of binaries due to mass ejection and kicks during the two supernovae. 
We implemented the binary survival probability after the first supernova explosion by modifying Equation~(\ref{eq:double_model}) using the results of \citet{kalogera96}. We used the final progenitor mass of \citet{zhang08} for the mass of the primary star before the explosion and various combinations of the initial and final mass for the secondary at the moment of the primary explosion. We also tried a number of relative kick velocities. Adding this to the calculation had no significant consequence.
We note, however, that the second supernova is more important for the survival of the system \citep[e.g.][]{dewi04,willemskalogera04,willems04,stairs06,wang06,wong10}. Finally, while it is believed the NS masses are little affected by mass transfer, this may not hold for the progenitor stars. However, we think that results of detailed binary evolution and population synthesis models can affect only the details of the DNS mass distribution and not the overall conclusions.

\section{Discussions \& Conclusions}
\label{sec:disc}

In this work, we present a Bayesian framework that directly compares the observed distribution of NS masses to the theoretical models of NS formation in supernovae. We illustrate this method on a sample of double neutron stars from \citet{ozel12} which did not experience significant mass accretion and should reflect the distribution of NS birth masses. We use the relation between the initial progenitor mass and the final remnant mass of \citet{zhang08}, tabulated for a range of explosion energies, primordial and solar compositions and for a momentum piston that drives the explosion positioned either at the base of the oxygen burning shell (the $S/\na = 4\,\kb$ models) or at the outer edge of the deleptonized core (the $\ye$ core models). We also investigate models with no fallback. We assume that the NS progenitors are independently drawn from a Salpeter IMF (independent mass model) or from a binary model where the primaries (progenitors of millisecond pulsars) are drawn from a Salpeter distribution and secondaries (progenitors of companions) are drawn from a distribution of mass ratios $P(q)$ that is either flat or strongly peaked at $q=1$.

We find a strong preference for binary models over models independently drawn from a Salpeter distribution (Fig.~\ref{fig:bayes_factor}). However, this is true only if we can correctly assign primary and secondary labels to the pulsars and their companions (especially to J1906$+$0746; Fig.~\ref{fig:mass_ratios}). Otherwise, the preference for binary models is weak. The mass distributions of primaries and secondaries are asymptotically consistent with the IMF, but the mass distribution of the primary progenitors is cut off at the minimum mass for NS production ($\mmin$), while for secondaries it is cut off at the maximum mass ($\mmax$), hence each DNS component probes different mean progenitor masses. We do not find any clear preference for either of the binary mass ratio distributions we considered. We do not find any preference for models with $\mmax$. We find that for a primordial composition locating the piston at $S/\na = 4\,\kb$ is strongly favored, while for solar composition models placing the piston at the $\ye$ core is more probable. Models with no fallback (or with strong explosions that produce little fallback) are always preferred because they minimize the width of the NS mass distributions (Figs.~\ref{fig:binary2d} and \ref{fig:binary2d_fallback}). Globally, the highest probability model has a uniform distribution for the binary mass ratios, solar composition, sets the remnant mass to that of the $\ye$ core, and has no fallback. This model also appears to be consistent with the data in an absolute ``frequentist'' sense (Figs.~\ref{fig:binary2d} and \ref{fig:ns_cumul}).

The preference for no or very little fallback could either be a feature of the supernova mechanism or a consequence of mass loss. Stripping the progenitor significantly reduces fallback, as can be seen in the high ($\ms \gtrsim 40\,\msun$) mass solar metallicity \citet{zhang08} models. For example, their model with an initial mass of $60\,\msun$ has pre-explosion mass of only $7.29\,\msun$ and mass inside $S/\na = 4\,\kb$ of $1.60\,\msun$, with a fallback mass of only $0.04\,\msun$ ($0.00\,\msun$) for $E = 1.2\times 10^{51}$\,ergs ($2.4\times 10^{51}$\,ergs). While the mass can be transferred between the stars in the binary and change the amount fallback when compared to a single star, the point is that our results prefer no fallback in either star of the binary. This is in agreement with the models of \citet{portegies98} where both of the two supernovae in a binary occur in a stripped progenitor. Alternatively, the DNS mass distribution may represent a surprisingly clear fingerprint of the supernova explosion mechanism. Specifically, with no fallback, the mass of the NS corresponds to the mass coordinate of the progenitor where the explosion was initiated. The bulk of the stellar mass growth and consequently also the bulk of the DNS population growth occurred between redshifts $1$ and $2$ \citep[e.g.][]{juneau05}, when the metallicity was approximately solar. Our $Z=Z_{\sun}$ results clearly prefer the edge of the deleptonized core ($S/\na \approx 2.8\,\kb$) as the mass coordinate of the explosion rather than at the base of the oxygen shell ($S/\na = 4\,\kb$).

\begin{figure}
\centering
\includegraphics[width=0.45\textwidth]{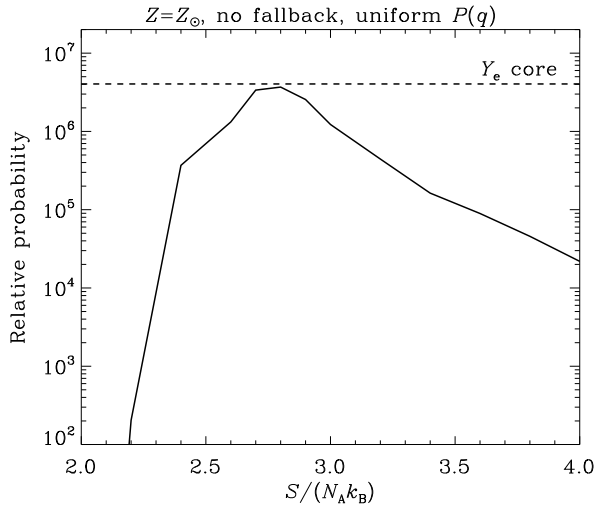}
\caption{Relative probability of the DNS mass models as a function of the position of the mass cut in entropy ($S/\na\kb$) that separates the remnant and the supernova ejecta (no fallback). The solid line shows the results for the supernova progenitors of \citet{woosley02}, while the dashed horizontal line is for the best model based on \citet{zhang08} that places the piston at the $\ye$ core and assumes no fallback (right panel of Fig.~\ref{fig:bayes_factor}).}
\label{fig:prog_cut}
\end{figure}

The two piston positions used by \citet{zhang08} are somewhat arbitrary so we explored whether other piston position would yield better results. We took the solar metallicity progenitor models of \citet{woosley02}\footnote{\url{http://www.stellarevolution.org/data.shtml}} that cover the mass range $10.8 \leq \ms \leq 40\,\msun$ and determined the mass coordinates that correspond to a range of values of the entropy, $2.0\,\kb \leq S/\na \leq 4.0\,\kb$. As we have no means of calculating the amount of fallback and models without fallback are preferred anyway, we assume that the remnant masses are set by the mass coordinate at the given entropy. The relative probabilities of these models were calculated with $\mmin$ and $M_{\rm ec}$ as free parameters and $\mmax$ fixed (Sec.~\ref{sec:res_limit}) and are shown as a function of the entropy in Figure~\ref{fig:prog_cut}. We see that there is a clear maximum at $S/\na \approx 2.8\,\kb$. The relative probability at maximum is almost equal to the highest probability model from Figure~\ref{fig:bayes_factor}, which is not surprising because for many progenitors this is the entropy at the edge of the $\ye$ core. If correct, this is an important result, because it shows that supernova explosions are initiated approximately at the edge of the iron core. Since the elapsed time from the core bounce until the accretion of all material out to the edge of the deleptonized core is $0.1$ to $0.3$\,s for majority of the progenitors, this also means that the supernovae explode by a delayed mechanism. That it is not longer challenges some of the delayed neutrino-driven models that explode at $\gtrsim 0.5$\,s after the bounce \citep{marek09,muller12} and constrains the time available for the unstable modes that potentially drive the explosion \citep{fryer12}. It is important to note that the composition and density of the layers that accrete through the shock do not change smoothly with time; the composition interfaces such as the one at the edge of the iron core lead to sudden decreases in the mass accretion rate, which cause expansion of the shock. This expansion might be temporary and can be followed by recession of the shock \citep{marek09}, or it might trigger an explosion such as those observed after the advection of the Si/SiO shell interface \citep{buras06a,muller12}. 

Independent of any concern about binaries, the greatest limitation of this and similar studies in the near future is our poor understanding of the physics of supernova explosions that link the initial progenitor mass to the final remnant mass. While the DNS mass distribution strongly suggests that the explosion develops at the edge of the deleptonized core, supernova theory does not provide any {\em a priori} reason why this should be so and whether this is true for all progenitor masses, metallicities, rotation rates and other parameters.

To summarize, ideal theoretical calculations of the massive star remnant mass function require an understanding of the supernova explosion mechanism. Until a self-consistent understanding is available, artificially induced supernova explosions intended to study the remnant population should increase their realism by manually inducing the explosions by either increasing the neutrino luminosity or absorption cross section in order to properly model the transition from accretion to neutrino-driven wind and the explosion\footnote{After the submission of this paper, \citet{ugliano12} published a NS mass distribution that is based on neutrino-driven explosions that are followed until the end of the fallback. We present the resulting probability distribution and comparison with other models in the Appendix.}, as has been done for some studies of explosion physics \citep[e.g.][]{scheck06,nordhaus10,nordhaus11,fujimoto11,hanke11,kotake11}. Although the DNS distribution suggests that fallback cannot be significant for the DNS systems, it would be interesting to see if this is also a feature of ``semi-physical'' explosion models. Of particular importance is whether fallback is a strong or stochastic function of progenitor structure, if it occurs. The range of progenitor models and model supernovae also needs to be expanded to better cover the progenitor mass range in order to fully sample the rapidly changing core properties with mass. Models are most needed near solar metallicities, since such stars overwhelmingly dominate any observable population of supernovae or NS binaries, and at the low masses that dominate the progenitor population due to the initial mass function. Observations mainly constrain the outcomes of low-mass solar metallicity stars - the models least examined in theoretical studies.

\section*{Acknowledgments}
This work is supported in part by an Alfred P. Sloan Foundation Fellowship and by NSF grant AST-0908816. We thank Kris Stanek for discussions and encouragement. We thank Thomas Janka for providing us with the NS mass distribution of \citet{ugliano12} and to the editor of our paper for allowing us to add the Appendix.

\begin{table*}
\centering
\caption{Summary of the remnant mass distribution models from \citet{zhang08}.}
\begin{tabular}{ccccc}
\hline
$Z$ & Piston at & $\mmin$ & Progenitor mass range &  $E$ $[10^{51}\,{\rm ergs}]$\\
\hline
$0$ & $S/\na=4\,\kb$ & $9.5\,\msun$ & $(10,100)\,\msun$ & 0.3, 0.6, 0.9, 1.2, 1.5, 1.8, 2.4, 3.0, 5.0, 10.0 \\
$0$ &  $S/\na=4\,\kb$ & $9.5\,\msun$ & $(10,100)\,\msun$ & $S/\na=4\,k_{\rm B}$ core mass, no fallback\\
$0$ & $\ye$ core & $9.5\,\msun$ & $(10,100)\,\msun$ & 1.2, 10.0 \\
$0$ & $\ye$ core& $9.5\,\msun$ & $(10,100)\,\msun$ & $\ye$ core mass, no fallback\\
$Z_{\sun}$ & $S/\na=4\,\kb$ & $9.1\,\msun$ & $(12,100)\,\msun$ & 1.2, 2.4 \\
$Z_{\sun}$ & $S/\na=4\,\kb$ & $9.1\,\msun$ & $(12,100)\,\msun$ & $S/\na=4\,k_{\rm B}$ core mass, no fallback\\
$Z_{\sun}$ & $\ye$ core & $9.1\,\msun$ & $(12,100)\,\msun$ & 1.2, 2.4 \\
$Z_{\sun}$ & $\ye$ core& $9.1\,\msun$ & $(12,100)\,\msun$ & $\ye$ core mass, no fallback\\
\hline
\end{tabular}
\label{tab:zhang}
\end{table*}

\begin{table*}
\begin{center}
\caption{Summary of double neutron star systems.}
\label{tab:dns}
\begin{tabular}{cllccllc}
\hline
Name & $\overline{M}_1$ & $\overline{\sigma}_1$ & $P_1$ [ms] & $\dot{P}_1$  & $\overline{M}_2$ & $\overline{\sigma}_2$ & Refererence \\
\hline
J0737-3039$^a$ & 1.3381 & 0.0007 & 22.7& $1.8\times 10^{-18}$ &  1.2489 & 0.0007 & \citet{kramer06}\\
B1534+12   & 1.3332 & 0.0010 & 37.9& $2.4\times 10^{-18}$ &  1.3452 & 0.0010 & \citet{stairs02}\\
J1756-2251 & 1.40   & 0.02   & 28.5& $1.0\times 10^{-18}$ &  1.18   & 0.02   & \citet{faulkner05}\\
J1906+0746 & 1.248  & 0.018  & 144.1&$2.0\times 10^{-14}$ &  1.365  & 0.018  & \citet{kasian08}, \citet{lorimer06}\\
B1913+16   & 1.4398 & 0.002  & 59.0& $8.6\times 10^{-18}$ &  1.3886 & 0.002  & \citet{weisberg10}\\
B2127+11C  & 1.358  & 0.010  & 30.5& $5.0\times 10^{-18}$ &  1.354  & 0.010  & \citet{jacoby06}\\
\hline
\end{tabular}
\end{center}
$^a$ The companion is also a pulsar with $P_2 = 2.8$\,s and $\dot{P}_2 = 8.9 \times 10^{-16}$.
\end{table*}

\appendix
\section{The NS mass distribution of Ugliano et al.\ (2012)}
\label{sec:ugliano}

\begin{figure*}
\centering
\includegraphics[width=0.8\textwidth]{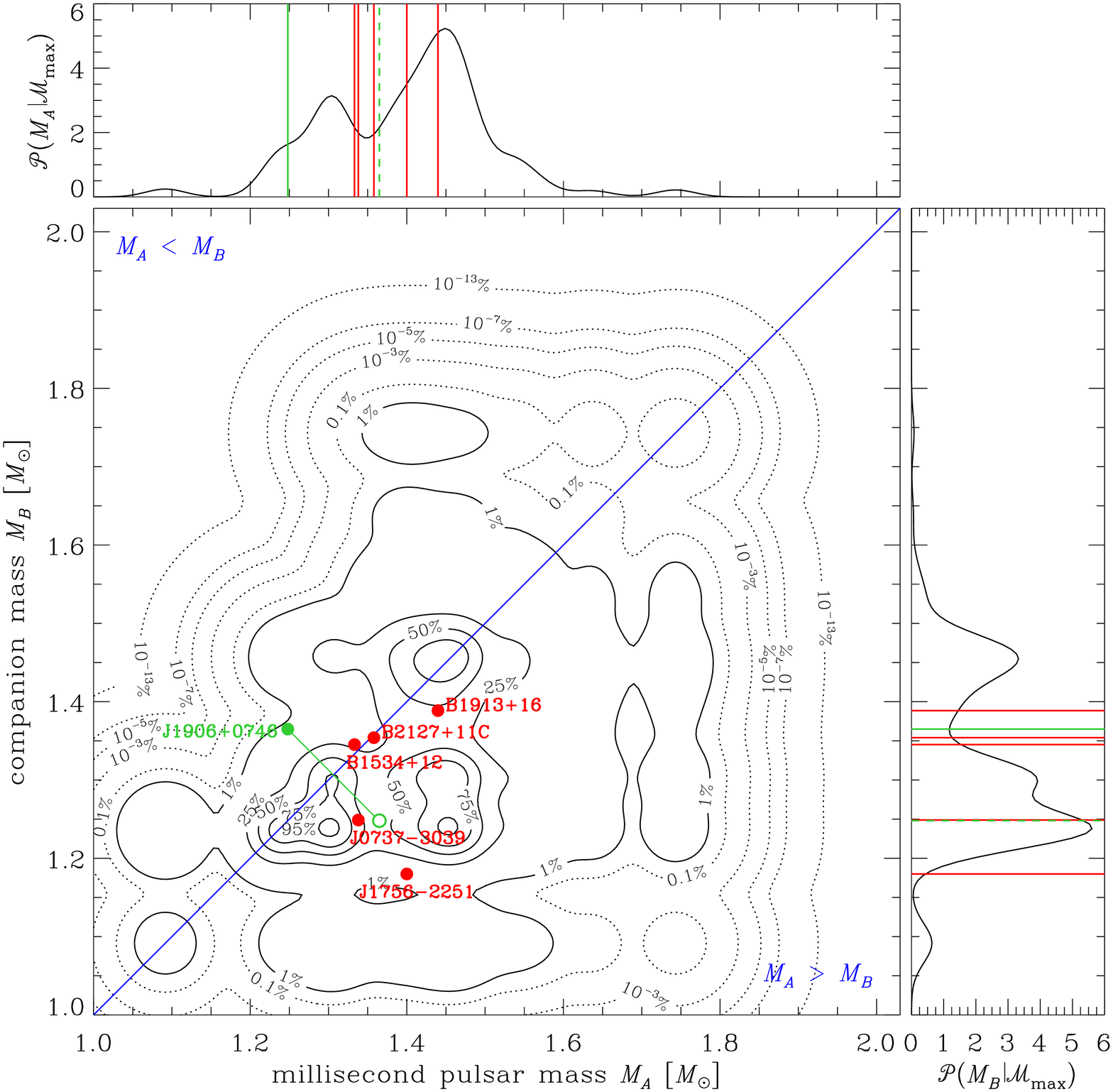}
\caption{Same as Figure~\ref{fig:binary2d}, but for a model based on the solar metallicity NS mass distribution of \citet{ugliano12}. The distribution of $P(q)$ is uniform and the calculation includes all progenitors of \citet{ugliano12}.}
\label{fig:binary2d_ugliano}
\end{figure*}

After the submission of this paper, \citet{ugliano12} published a NS mass distribution that is based on 1D simulations of neutrino-driven explosions followed from the onset of the collapse until the end of the fallback. The simulations are normalized by comparison with the observed parameters of SN 1987A. The distribution is based on over $100$ progenitors with {$10\,\msun \leq \ms \leq 40\,\msun$}, which we supplement with remnant masses of $1.35\,\msun$ for $9.1\,\msun \leq \ms < 10\,\msun$, similarly to the models of \citet{zhang08} discussed in the main paper. We also corrected the remnant masses for the binding energy according to Equation~(\ref{eq:grav}). In Figure~\ref{fig:binary2d_ugliano}, we show the resulting probability distribution of the DNS masses. We see that the peaks of the distribution are shifted by $\sim 0.1\,\msun$ to higher masses with the respect to the DNS data. Even though the fallback on the remnants is included, the probability distribution has little power for NS masses higher than about $1.6\,\msun$, unlike the large amounts of fallback in many of the \citet{zhang08} models (Fig.~\ref{fig:binary2d_fallback}). Indeed, \citet{ugliano12} find little fallback for high-mass progenitors.

In order to compare the relative probability of the \citet{ugliano12} distribution to the models in Figure~\ref{fig:prog_cut}, we marginalize over $\mmin$ and the remnant mass $M_{\rm ec}$ for progenitors with $\mmin \leq \ms < 10.8\,\msun$ in the same way as we did for the models of \citet{woosley02}. We find that the relative probability is only $10^{3.2}$ on the scale of Figure~\ref{fig:prog_cut}, which is likely caused by the slight mismatch in the probability peaks with respect to the DNS data. If we include remnant masses without fallback, which might be more appropriate for DNS progenitors with stripped hydrogen envelopes, the relative probability raises to about $10^{5.4}$. However, this is about a factor of $10$ worse than our best models in Figures~\ref{fig:bayes_factor} and \ref{fig:prog_cut}. Thus, the reduction of fallback in the \citet{ugliano12} NS mass distribution provides a better match to the observed masses of DNS binaries, but not as good match as the \citet{zhang08} $\ye$ core models with no fallback.

\end{document}